\documentstyle[12pt]{article}
\makeatletter
\newcommand{\singlespacing}{\let\CS=\@currsize\renewcommand{\baselinestretch}
{1.0}\tiny\CS}
\newcommand{\doublespacing}{\let\CS=\@currsize\renewcommand{\baselinestretch}
{1.5}\tiny\CS}
\begin{document}
\textwidth 16cm
\newcommand{\bd}{\begin{document}}
\newcommand{\ed}{\end{document}}
\newcommand{\bc}{\begin{center}}
\newcommand{\ec}{\end{center}}
\newcommand{\bfr}{\begin{flushright}}
\newcommand{\efr}{\end{flushright}}
\newcommand{\lt}{\left}
\newcommand{\rt}{\right}
\newcommand{\vs}{\vspace}
\newcommand{\hs}{\hspace}
\newcommand{\beq}{\begin{equation}}
\newcommand{\eeq}{\end{equation}}
\newcommand{\lb}{\linebreak}
\renewcommand{\pb}{\pagebreak}
\newcommand{\mb}{\makebox}
\newcommand{\fb}{\framebox}
\newcommand{\mc}{\multicolumn}
\newcommand{\ben}{\begin{enumerate}}
\newcommand{\een}{\end{enumerate}}
\newcommand{\bit}{\begin{itemize}}
\newcommand{\eit}{\end{itemize}}
\newcommand{\ol}{\overline}
\newcommand{\un}{\underline}
\newcommand{\lefq}{\lefteqn}
\newcommand{\ba}{\begin{array}}
\newcommand{\ea}{\end{array}}
\newcommand{\beqa}{\begin{eqnarray}}
\newcommand{\eeqa}{\end{eqnarray}}
\newcommand{\beqas}{\begin{eqnarray*}}
\newcommand{\eeqas}{\end{eqnarray*}}
\newcommand{\bfg}{\begin{figure}}
\newcommand{\efg}{\end{figure}}
\newcommand{\bds}{\begin{displaymath}}
\newcommand{\eds}{\end{displaymath}}
\newcommand{\btb}{\begin{tabbing}}
\newcommand{\etb}{\end{tabbing}}
\newcommand{\para}{\parallel}
\newcommand{\pad}{\partial}
\newcommand{\nn}{\nonumber}
\newcommand{\la}{\leftarrow}
\newcommand{\ra}{\rightarrow}
\newcommand{\lgla}{\longleftarrow}
\newcommand{\lgra}{\longrightarrow}
\newcommand{\La}{\Leftarrow}
\newcommand{\Ra}{\Rightarrow}
\newcommand{\Lra}{\Leftrightarrow}
\newcommand{\Lgla}{\Longleftarrow}
\newcommand{\Lgra}{\Longrightarrow}
\newcommand{\bm}{\boldmath}
\newcommand{\lan}{\langle}
\newcommand{\ran}{\rangle}
\renewcommand{\a}{\alpha}
\renewcommand{\b}{\beta}
\newcommand{\g}{\gamma}
\newcommand{\G}{\Gamma}
\renewcommand{\d}{\delta}
\newcommand{\eps}{\epsilon}
\newcommand{\th}{\theta}
\newcommand{\Th}{\Theta}
\newcommand{\s}{\sigma}
\newcommand{\lam}{\lambda}
\newcommand{\D}{\Delta}
\newcommand{\vare}{\varepsilon}
\newcommand{\pr}{\prime}
\newcommand{\ro}{\rho}
\newcommand{\nab}{\nabla}
\newcommand{\m}{\mu}
\newcommand{\n}{\nu}
\newcommand{\Sg}{\Sigma}
\newcommand{\p}{\pi}
\newcommand{\R}{I\!\!R}
\newcommand{\om}{\omega}
\newcommand{\Om}{\Omega}
\newcommand{\ze}{\zeta}
\newcommand{\vart}{\vartheta}
\newcommand{\tri}{\triangle}
\newcommand{\f}{\frac}
\newcommand{\iny}{\infty}
\newcommand{\pro}{\propto}

\centerline{\Large \bf ENERGY SPECTRA OF TWO ELECTRONS }

\vs{1cm}

\centerline{\Large \bf IN A CIRCULAR QUANTUM DOT}

\vs{2cm}

\centerline{ ANJANA SINHA$^*$}
\centerline{\it Department of Applied Mathematics,}
\centerline{\it University of Calcutta,}
\centerline{\it 92, A.P.C. Road, Kolkata - 700 009, INDIA}

\vs{.5cm}

\centerline{\it and }

\vs{.5cm}

\centerline{    Y.P. VARSHNI $^\$ $}
\centerline{\it Department of Physics,}
\centerline{\it University of Ottawa,}
\centerline{\it Ottawa, CANADA  K1N 6N5}

\vs{2cm}

---------------------------------------------------------
-----------------------------------------\\

\noindent
e-mail: \\
a.sinha@cucc.ernet.in, anjana23@rediffmail.com \\
\$ ypvsj@aix1.uottawa.ca

\pb

\noindent
{\large \bf \un{ABSTRACT :}}

\vs{1cm}

The electron interaction energy of two interacting electrons 
in a  circular quantum dot (with hard wall confinement )
                  is investigated in the framework of the semi-classical
                  Wentzel-Kramers-Brillouin (WKB) approximation.
            The two electrons are assumed to be in an infinitely deep well 
                  of radius
                  $r_0$, in a simple configuration with one electron fixed at 
                  the origin.
                  The corresponding Schr\"{o}dinger equation, with hard wall
                  boundary conditions, is also solved exactly by numerical 
                  integration. It is
                  observed that the agreement between the two energy values is
                  quite good, suggesting that the WKB approximation works well 
                  for such a
                  confined quantum system as well. This may provide motivation 
                  to extend this to
                  more realistic confined potentials.

                  \pb

                  \noindent
                  {\Large \bf 1. Introduction  }
                  
\vs{0.5cm}

                  Advances in submicron technology have made it possible to
                  realize artificial nanostructures, where electrons can be 
                  confined on a scale
                  comparable to their de Broglie wavelengths. These
                  quantum dots, or {\it artificial atoms} as they are called,
                  containing only a few electrons, exhibit unusual optical and 
                  transport
                  properties, because of the discrete energy levels of the 
                  confined electrons.
                  $^{1}$. Quantum dots have been extensively
                  investigated both experimentally and theoretically$^{1-4}$.
                  Ever since the integral and fractional quantum Hall effects
                  were discovered in two-dimensional electron systems
                  under a high magnetic field, two-dimensional
                  studies have been the subject of intense investigation in 
                  quantum solid state
                  physics and quantum field theory. In two-dimensional quantum 
                  dots the
                  electrons are confined in a circular region, such that the 
                  motion
                  perpendicular to the plane of confinement is essentially 
                  frozen out.
                  Such quantum dots can be fabricated with two types of 
                  confinement potential
                  --- the `soft' confinement potential, accomplished by 
                  electric and / or
                  magnetic fields, or a `hard' confinement in pillar-shaped
                  quantum dots.

                  To study the behaviour of single electron tunneling, it is 
                  essential to derive
                  the eigensolutions of the system of interacting electrons 
                  confined in an
                  ultrasmall quantum dot. Different theoretical methods have 
                  been used to
                  study the problem of two
                  electrons in a two-dimensional quantum dot, both with a soft 
                  wall (parabolic
                  confinement) as well as a hard wall (rigid circular wall at a 
                  radius $r_0$)
                  in-plane confinement. Physical examples of electrons being
                  confined in such a thin layer can be found in the
                  vicinity of junctions between insulators and semiconductors, 
                  between layers of
                  semiconductors, and between a vacuum and liquid helium. 
                  Consequently, this
                  field of research is of vital importance to modern electronic 
                  devices.

                  Merkt, Huser and Wagner$^{5}$ have calculated the energy 
                  spectra of two
                  electrons in a two-dimensional harmonic quantum dot, in the 
                  effective mass
                  approximation, as a function of the dot size and the strength 
                  of a magnetic
                  field directed perpendicularly to the dot plane. Matulis and 
                  Peeters$^{6}$ have
                  proposed a convergent renormalized perturbation series in 
                  powers of the
                  electron-electron interaction for calculating the energy of a 
                  quantum dot.
                  They have used this method to calculate the ground and several 
                  excited states
                  of a quantum dot consisting of two electrons. Matulis, 
                  Fjaerestad and Chao$^{7}$
                  have solved the Schr\"{o}dinger equation for the ground state 
                  of two electrons
                  in a two-dimensional circular quantum dot, with hard 
                  confinement potential,
                  using a renormalized perturbation series approach, which
                  interpolates between the perturbation solutions in the weak 
                  interaction regime
                  and the asymptotic solutions in the strong interaction regime. 
                  They assumed
                  that one electron was fixed at the origin. Zhu et al$^{8}$ 
                  have made use of the
                  expansion in a power series to obtain the eigen solutions of 
                  two electrons in
                  a parabolic quantum dot. McKinney and Watson$^{9}$ have 
                  applied the dimensional
                  perturbation theory to the two-electron D-dimensional quantum 
                  dot, obtaining
                  values for the ground- and excited-state energies. The charge- 
                  and
                  spin-density excitation spectra for two electrons in a 
                  two-dimensional
                  circular hard-wall confined quantum dot have been calculated 
                  by Brataas, Hanke
                  and Chao$^{10}$. Akman and Tomak$^{11}$ have performed the 
                  exact numerical
                  diagonalization of the Hamiltonian of a 2D circular quantum 
                  dot for 2, 3, and
                  4 interacting electrons. Adamowski et al$^{12}$ have studied 
                  two electrons
                  confined in quantum dots under an assumption of a Gaussian 
                  confining potential
                  and its parabolic approximation.  They have calculated the 
                  energy levels of
                  singlet and triplet states as functions of the range and depth 
                  of the
                  confining potential in the two-dimensional (circular) and 
                  three-dimensional
                  (spherical) quantum dots. There have been several other 
                  investigations on the
                  properties of circular quantum dots.

                  In the present work, the energy spectrum of two interacting
                  electrons confined in a rigid disc of radius $r_0$, is 
                  investigated
                  within the effective mass
                  approximation, using the WKB method. The dot is considered in 
                  the
                  two-dimensional limit of thin discs. The simplest model of 
                  quantum
                  confinement is assumed, viz., infinitely deep well of radius 
                  $r_0$.

                  \vs{2cm}

                  \noindent
                  {\Large \bf 2. Theory  }
                  
\vs{.5cm}

                  We apply the WKB method to the case of two particles,
                  confined in a thin, circular, two-dimensional disc of radius 
                  $r_0$,
                  lying in the $xy$ plane. The Hamiltonian for such a system is 
                  then given by
                  $$ H = - \f{\hbar^2  }{ 2\mu ^*} \nabla _1 ^2 -  \f {\hbar^2 
                  }{  2\mu ^*}
                  \nabla _2 ^2 + V(r_1) +V(r_2) + V(|r_1 - r_2|)
                  \eqno (1) $$
                  where $ \mu ^* $ is the effective mass. In two dimensions,
                  the radial distance $r$ and the angle $\theta$ are given by
                  $$ r = \sqrt{ x ^2 + y ^2 } , \ \ \ \ \ \theta  = tan ^{-1}
                  \left ( \f{y}{x} \right ) \eqno (2) $$

                  We shall consider here the same configuration as that 
                  considered by
                  Matulis, Fjaerestad and Chao$^{7}$, that is
                       one electron is fixed at the centre of the disc-like 
                  quantum dot.
                  Substituting the radial wavefunction $ \psi (r)$ by
                  $$ \psi (r) = \f{u(r)}{\sqrt{r}} \eqno (3) $$
                  the radial Schr\"odinger equation can be written as an
                  effective one-dimensional equation
                  $$  \f{d^2 u(r)}{dr^2} + \Gamma ^2 (r) u(r) = 0 \eqno (4) 
$$
                  where
                  $$ \Gamma ^2 (r) = \f{2 \mu ^*}{\hbar^2}
                  \left\{ E -  \f{1}{r} -  \f{(m ^2 -1/4) \hbar^2}{2 \mu 
                  ^* r^2 } \right\} \eqno (5) $$
                  and $m$ stands for the azimuthal quantum number. The Coulomb 
                  repulsion
                  term $  \f{1}{r} $ is responsible for the rich
                  structure of the energy spectrum of the quantum dot.
                  To apply  the semiclassical Wentzel-Kramers-Brillouin (WKB)
                  approximation, the $0 < r < \infty $ space is conformally 
                  mapped to the
                  $-\infty < w  < \infty$ space, by substituting$^{13}$
                  $$ r  = e^{w } \eqno (6) $$
                  This mapping prevents the left turning point to be negative  
                  for attractive potentials (for the case $m = 0$).

                  Thus the Schr\"odinger eqn. (4) gets transformed to
                  $$   \f{d^2 \chi (w)}{dw^2} + \Gamma^2 (w) \chi(w) = 0 
\eqno (7) $$
                  where
                  $$ \Gamma^2 (w) = \left\{ \f{2\mu ^*}{\hbar^2} e^{2w}
                  \left[ E - V(w)\right] - m ^2 \right\} \eqno (8) $$
                  with
                  $$   \chi (w) =   \psi(r)
                  =   e^{-w/2} u(w) \eqno (9) $$
                  In terms of the original variable $r$,
                  $$ \int^w \Gamma (w) dw = \int^r \Gamma (r) dr \eqno (10) $$
                  where
                  $$ \Gamma^2 (r) = \f{2\mu ^*}{\hbar^2} \left[ E - V_1 
                  (r) \right]
                  \eqno (11) $$
                  and
                  $$ V_1 (r) = V(r) + \f{m ^2 \hbar^2}{2 \mu ^* r^2} \ \ 
                  \ \ \ \
                  m = 0, \pm 1, \pm 2, \ldots \eqno (12) $$

                  Thus in two dimensions, conformal mapping replaces $(m ^2 - 
                  1/4)$
                  by $m ^2$ (Yi et al$^{13}$). In other words, the centrifugal
                  barrier term $\f{(m ^2 - 1/4) \hbar^2}{2\mu r^2}$ is 
                  modified
                  to $ \f{m ^2 \hbar^2}{2\mu r^2}$. The situation might
                  be compared with Langer modification in three dimensions, 
                  where the
                  centrifugal term $ \f{l(l+1)\hbar^2}{2\mu r^2}$ is 
                  replaced by
                  $\f{l(l+1/2)\hbar^2 }{ 2\mu r^2}$.
                  Henceforth, for convenience of calculations, the units used 
are
                  $\hbar = 1$.

                  Assuming an infinite deep well model, the  hard wall 
                  confinement condition
                  implies
                  $$ V(r) = \infty ~~~~~~~ r > r_0 $$
                  $$ V(r) = 0 ~~~~~~~~ r < r_0  \eqno (13) $$
                  The radial solution $u(r)$ satisfies the Schr\"{o}dinger
                  equation,
                  $$ - \f{ 1}{ 2 \mu ^* } \f{d^2 u }{ dr^2} +
                  \left \{ \f{1}{r} + \f{ m ^2}{2 \mu ^* r^2 } \right 
                  \} u  = E u \eqno (14) $$
                  Substituting
                  $$ \rho = \mu ^* r  \eqno (15) $$
                  $$ e = \f{ E}{ \mu ^*} \eqno (16) $$
                  equation (14) reduces to
                  $$  \f{d^2 u}{ d \rho ^2} +  \Gamma ^2 (\rho) u  = 0 \eqno 
                  (17) $$
                  where
                  $$ \Gamma ^2 (\rho) = 2 \left \{ e - V \right \} \eqno (18) $$
                  $$ V = \f{1}{ \rho} +  \f{m ^2}{2 \rho ^2} \eqno (19) 
                  $$
                  The impenetrable circular wall imposes the
                  additional boundary condition
                  $$ u(r_0) = 0 \eqno (20) $$

                  \noindent
                  It may be mentioned here that the region of space
                  $0 \leq \rho \leq r_0$ may be divided
                  into 2 sections

                  \noindent
                  i) Region I : $ 0 \leq \rho \leq \rho _t $ where $V > e$

                  \noindent
                  ii) Region II : $ \rho _t \leq \rho < r_0 $ where $e > V$

                  \noindent
                  where $\rho _t $ is the classical turning point, obtained by 
                  putting
                  $\Gamma ^2 (\rho)   = 0$.

                  A WKB ansatz is assumed for the solution of the
                  Schr\"{o}dinger equation  in region I, and
                  the solution in region II  is obtained with the
                  help of the WKB connection formulae$^{14}$ on either side
                  of the turning point as,
                  $$ u_{I} (\rho) = \f{A}{ \sqrt{\kappa (\rho)}} \exp 
\left\{
                  - \int^{\rho_t } _{\rho} {\kappa}  d \rho \right\} \eqno (21) 
$$
                  $$ u_{II} (\rho) = \f{2A}{ \sqrt {\Gamma (\rho)}}
                  \sin \left\{ \int^{\rho}_{\rho _t} \Gamma (\rho) d \rho
                  + \f{\pi}{4} \right\} \eqno (22) $$
                  where
                  $$ \kappa ^2 (\rho)  = - \Gamma ^2 (\rho)  \eqno (23) $$

                  Imposition of the condition $u_{II} (r_0) = 0$
                  gives the WKB quantization rules as
                  $$ \alpha  = \left( n_r + \f{3}{4} \right) \pi ~~~~~~~
                  n_r = 0, 1, 2, \cdots \eqno (24) $$
                  where $ \alpha $ is calculated to be$^{15}$
                  $$ \alpha  = \int^{r_0}_{\rho _t} \left[ 2 \left\{ e - 
                  \f{1}{ \rho} -
                  \f{m ^2}{2 \rho ^2 } \right\} \right]^{1/2} d \rho $$
                  $$ = \sqrt{2e r_0 ^2 - 2 r_0 - m ^2}
                  + |m|  sin ^{-1}  \f{ r_0 + m ^2}{ r_0 \sqrt{1 + 2 e m ^2} }
                  -  |m| \f{\pi}{2} $$
                  $$ -  \f{1}{ \sqrt{2 e}}
                  ln \left\vert  { \sqrt{2e r_{00}} + 2e r_0 -1}
                  {2e \rho _t -1 } \right\vert  \eqno (25) $$
                  with
                  $$ \rho _t =  \f{ 1 + \sqrt{ 1 + 2 e m ^2}}{2 e }  
                  \eqno(26) $$
                  $$ r_{00} = 2 e r_0 ^2 - 2 r_0 - m ^2 \eqno (27) $$

                  \vs{1cm}

                  \noindent
	            {\Large \bf 3. Results and Discussion }

                  \vs{.5cm}

                  The WKB energies $E (WKB)$ for the system were calculated for 
                  various values of
                  the confining radius, with the help of the equations given 
                  above.
                  The exact energies $E (exact)$ are obtained by numerical 
                  integration of the
                  Schr\"{o}dinger equation using Numerov's method and the 
                  logarithmic mesh.
                  The effective mass has been taken as $ \mu ^* =  \f{1}{4} 
$.
                  To test the validity of our approach, our WKB results are 
                  compared
                  with the exact numerical energies
                  for different values of the confining radius $r_0$, in Table 1
                  ($n_r = 0, m = 0$),  Table 2
                  ($n_r = 0, m = 1$),  and Table 3 ($n_r = 1, m = 1$).
                  Considering the semiclassical nature of the WKB approximation, 
                  the
                  agreement between the two values is found to be excellent, 
                  improving further
                  as the region of confinement increases. It may be mentioned 
                  here that though
                  the infinitely deep well circular quantum dot with two 
                  electrons was
                  investigated by Goff and St\'{e}b\'{e}$^{16}$ and Matulis, 
                  Fjaerestad and
                  Chao$^{7}$,
                  their discussions were restricted to the ground state only, 
                  whereas the
                  analysis presented in this paper is valid for excited states 
                  as well.
                  Moreover, the approach used here is totally different from 
                  either of these
                  cases.

                  To conclude the confined energies are obtained for two 
                  interacting electrons
                  in a circular quantum dot of radius $r_0$, with one electron 
                  fixed at the
                  origin. The semiclassical WKB approach adopted here
                  gives excellent results when compared with exact numerical 
                  values. In this
                  work the simplest model of the confined two-electron
                  system is considered, by assuming an infinitely deep well for 
                  the
                  confining potential. The authors hope to deal with more
                  complicated potentials in future.

			\pb

\noindent
Table 1. Energy of the ground state ($n_r = 0, m = 0$)

\vs{0.5cm}

\begin{tabular}{|c|c|c|} \hline
$r_0$ & $E(WKB)$  & $E(exact)$ \\ \hline
0.4 & 47.031   & 44.505 \\ \hline
0.6 & 22.989   & 21.589 \\ \hline
0.7 & 17.608   & 16.513 \\ \hline
0.8 & 14.013   & 13.135 \\ \hline
0.9 & 11.479   & 10.762 \\ \hline
1.0 & 9.6186   & 9.0240   \\ \hline
1.5 & 4.9446   & 4.6693   \\ \hline
2.0 & 3.1283   & 2.9776   \\ \hline
3.0 & 1.6742   & 1.6152   \\ \hline
4.0 & 1.0895   & 1.0610   \\ \hline
5.0 & 0.78678  & 0.7712   \\ \hline
6.0 & 0.60592  & 0.59663  \\ \hline
9.0 & 0.34399  & 0.34126  \\ \hline
10 &  0.29788  & 0.29592  \\ \hline
12 &  0.23288  & 0.23180  \\ \hline
\end{tabular}

\vs{2cm}

\noindent
Table 2. Energy of the excited state having $n_r = 0, m = 1$.

\vs{0.5cm}

\begin{tabular}{|c|c|c|} \hline                  
$r_0$ & $E(WKB)$  & $E(exact)$ \\ \hline
0.5  &  65.835   &   66.613 \\ \hline
1.0  &  18.479   &  18.646  \\ \hline
1.5  &   9.0618  &  9.1573   \\ \hline
2.0  &   5.6020   &  5.6328   \\ \hline
3.0  &   2.9121   &  2.9225  \\ \hline
5.0  &   1.3404   &  1.3428  \\ \hline
6.0  &   1.0288   &  1.0303  \\ \hline
8.0  &   0.68606  &  0.68676  \\ \hline
10   &   0.50576  &  0.50621   \\ \hline
15   &   0.29604  &  0.29629   \\ \hline
20   &   0.20500  &  0.20518 \\ \hline
\end{tabular}
                  
\pb
             
\noindent
Table 3. Energy of the excited state having ($n_r = 1, m = 1$).

\vs{0.5cm}

\begin{tabular}{|c|c|c|} \hline
$r_0$ & $E(WKB)$  & $E(exact)$ \\ \hline
0.5  &   206.42  &   206.51   \\ \hline
1.0  &   54.222  &   54.208   \\ \hline
1.2  &   38.375  &  38.356 \\ \hline
1.5  &   25.249  &   25.227  \\ \hline
2.0  &   14.842  &  14.821  \\ \hline
3.0  &   7.1557  &   7.1399  \\ \hline
4.0  &   4.3329  &   4.3209  \\ \hline
5.0  &   2.9659  &   2.9567  \\ \hline
6.0  &   2.1910  &   2.1837  \\ \hline
8.0  &   1.3763  &   1.3714  \\ \hline
10   &   0.96993 &   0.96643 \\ \hline
15   &   0.52556 &   0.52379  \\ \hline
20   &   0.34602 &   0.34496  \\ \hline
\end{tabular}

\vs{2cm}                  

\noindent
{\large \bf  Acknowledgment }

\vs{0.5cm}

                  One of the authors (A.S.) is grateful to CSIR, Govt. of India 
                  for
                  providing her with a Senior Research Associateship
{\it (Scientists' Pool Scheme)}.
                  This work was also supported in part by a research grant from 
                  the Natural
                  Sciences and engineering Research Council of Canada to the 
                  other author
                  (Y.P.V.).

                  \pb

                  \noindent
                  {\bf References}

                  \vs{0.5cm}

			\begin{enumerate}
                  \item[1.] M. A. Reed and W. P. Kirk (editors), Nanostructure 
                  Physics and
                  Fabrication (Academic, Boston, 1989).
                  \item[2.] L. Jacak, P. Hawrylak and A. W\'{o}js, Quantum Dots
                   (Springer-Verlag, 1997).
                  \item[3.] D. Bimberg, M. Grundmann and N. N. Ledentsov, 
                  Quantum Dot
                  Heterostructures  (John-Wiley, 1998).
                  \item[4.] A. D. Yoffe, Adv. Phys. {\bf 50}, 1 (2001).
                  \item[5.] U. Merkt, J. Huser, and M. Wagner, Phys. Rev. B 
                  {\bf 43},
                  7320 (1991).
                  \item[6.] A. Matulis and F. M. Peeters, J. Phys. Condensed 
                  Matter {\bf 6},
                  7751 (1994).
                  \item[7.] A. Matulis, J. O. Fjaerestad and K. A. Chao, Int. 
                  J. Mod. Phys. B
                  {\bf 11}, 1035 (1997).
                  \item[8.] Jin-Lin Zhu, Jing-Zhi Yu, Zhi-Qiang Li and Y. 
                  Kawazoe,
                   J. Phys. Condensed Matter  {\bf 8}, 7857 (1996).
                  \item[9.] B. A. McKinney and D. K. Watson, Phys. Rev. B {\bf 
                  61}, 4958 (2000).
                  \item[10.] A. Brataas, U. Hanke and K. A. Chao, Phys. Rev. B 
                  {\bf 54},
                  10736 (1996).
                  \item[11.] N. Akman and M. Tomak, Physica B {\bf 262}, 317 
                  (1999).
                  \item[12.] J. Adamowski, M. Sobkowicz, B. Szafran and S. 
                  Bednarek, Phys. Rev.
                   B {\bf 62}, 4234 (2000).
                  \item[13.] H. S. Yi, H. R. Lee and K. S. Sohn, Phys. Rev. A
                  {\bf 49}, 3277 (1994).
                  \item[14.] A. K. Ghatak, R. L. Gallawa  and I. C. Goyal,
                  MAF and WKB solutions to the wave equations. NIST Monograph 
                  176 (NIST, Washington
                  D.C., 1991).
                  \item[15.] M. Abramowitz and I. A. Stegun, Handbook of 
                  Mathematical Functions
                  (Dover Publications, Inc. New York, 1970)
                  \item[16.] S. L. Goff and B St\'{e}b\'{e}, J.Phys. B : At. 
                  Mol. Opt. {\bf 25}
                  5261 (1992).
			\end{enumerate}

                  \pb

\end{document}